# V2X Communication Between Connected and Automated Vehicles (CAVs) and Unmanned Aerial Vehicles (UAVs)

Ozgenur Kavas-Torris, Sukru Yaren Gelbal, Mustafa Ridvan Cantas, Bilin Aksun-Guvenc, Levent Guvenc

Department of Mechanical and Aerospace Engineering, The Ohio State University, Automated Driving Lab (ADL), Columbus, OH 43212 USA

Corresponding author: Ozgenur Kavas-Torris (e-mail: kavastorris.1@osu.edu)

**ABSTRACT** Connectivity between ground vehicles can be utilized and expanded to include aerial vehicles for coordinated missions. Using Vehicle-to-Everything (V2X) communication technologies, a communication link can be established between Connected and Autonomous vehicles (CAVs) and Unmanned Aerial vehicles (UAVs). Hardware implementation and testing of a ground to air communication link is crucial for real-life applications. Two different communication links were established, Dedicated Short Range communication (DSRC) and 4G internet based WebSocket communication. Both links were tested separately both for stationary and dynamic test cases. One step further, both links were used together for a real-life use case scenario called Quick Clear demonstration. The aim was to first send ground vehicle location information from the CAV to the UAV through DSRC communication. On the UAV side, the connection between the DSRC modem and Raspberry Pi companion computer was established through User Datagram Protocol (UDP) to get the CAV location information to the companion computer. Raspberry Pi handles 2 different connection, it first connects to a traffic contingency management system (CMP) through Transmission Control Protocol (TCP) to send CAV and UAV location information to the CMP. Secondly, Raspberry Pi uses a WebSocket communication to connect to a web server to send photos taken by an on-board camera the UAV has. Quick Clear demo was conducted both for stationary test and dynamic flight tests. The results show that this communication structure can be utilized for real-life scenarios.

## I. INTRODUCTION

Connectivity between ground vehicles have accelerated the research and development in ground vehicle-based intelligent transportation systems and applications. Connected and Autonomous vehicles (CAVs) can communicate with roadway infrastructure around them through Vehicle-to-Infrastructure (V2I) communication, which brings about traffic light signal and timing (SPaT) based vehicle speed planning to save fuel and improve mobility. CAVs are also able to communicate with each other through Vehicle-to-Vehicle (V2V) communication. Through V2V communication, CAVs can share their position, speed and acceleration information with other CAVs around them. Using the nearby vehicle information, CAVs can execute coordinated moves such as forming platoon and convoys to save fuel. CAVs can also communicate with other traffic agents such as pedestrians through Vehicle-to-Everything (V2X) communication. Using V2X, CAVs can be utilized in preventing unwanted collisions and accidents with pedestrians and bicyclists, as well as improving overall safety for other ground transportation agents.

Connectivity has been utilized widely for fuel economy improvement in CAVs in literature. Yu *et al.* [1] investigated fuel economy in ecological driving of individual and platooning vehicles using longitudinal autonomy and connectivity. Altan *et al.* [2] modeled a V2I algorithm for longitudinal control of a CAV to get smooth acceleration and deceleration speed profiles using SPaT information from an upcoming traffic light. Sun *et al.* [3] investigated different fuel-optimal methods for speed planning of CAVs. Asadi and Vahidi [4] devised a V2I Model Predictive Controller (MPC) that uses upcoming traffic light information to obtain both fuel savings and a shorter trip time. Similarly, Yu *et al.* [5] designed an MPC for eco-driving, however, their focus was on a platoon rather than a single vehicle. Xu *et al.* [6] considered Eco-driving for a transit rather than a personal vehicle to conserve fuel while reducing undesired emissions. Research has also been conducted on developing adaptive strategies for a dynamic roadway traffic environment while focusing on fuel savings [7].

Simulation tools that enable researchers to test CAV based algorithms are crucial for modelling and simulating algorithms that include connectivity [8]. Other than small scale implementation of connectivity technologies for a single CAV, the effects of having varying levels of CAV in a heterogeneous traffic flow has also been studied for future implementation of large scale deployment of CAVs on public roadways [9]- [10].

Recent advancements in Unmanned Aerial Vehicle (UAVs) technology have brought about a wide range of areas where UAVs are utilized for intelligent transportation system applications. UAVs have been used in flight missions for search and rescue operations [11], as well as out-of-sight indoor flights with tactile feedback [12]. UAVs have also been utilized in transportation of goods, parcels and passengers [13]. In agriculture, UAVs can be used to monitor the height and health of crops using on-board cameras.

Capabilities of UAV can be expanded by introducing intervehicular communication. UAVs can be equipped with communication links to establish UAV to UAV or UAV to CAV communication. Dedicated Short Range Communication (DSRC) can be used to setup a communication link between UAVs and CAVs. Currently, Automatic Dependent Surveillance-Broadcast (ADS-B) is used as the standard protocol to transmit location information in the aerospace industry. However, Moore *et al.* [14] has suggested that ADS-B will not be able to handle the low-altitude air traffic communication soon due to the expected increase in air traffic density at low altitude. Chakrabarty *et al.* [15] investigated how DSRC can be used as an alternative to ADS-B for UAV to UAV communication to prevent mid-air collisions at low altitudes. Menouar *et al.* [16] studied the applicability and challenges of how a UAV-enabled Intelligent Transportation System (ITS) for a smart city.

UAVs can also be utilized in coordinated missions with ground vehicles. Kavas-Torris *et al.* [17] developed use case scenarios to simulate CAV and UAV coordination, as well as demonstrated the hardware implementation for a DSRC based and a 4G WebSocket based CAV and UAV communication link.

In this paper, hardware implementation and real-life testing of a coordinated CAV & UAV mission is presented. Methodology section starts by introducing details about the CAV platform and the UAV platform used for the hardware implementation. Then, the software platform used for the real-life testing, the Contingency Management Platform (CMP) that receives information from the CAV, as well as the companion on-board computer Raspberry Pi 4B and the Webserver that was used to display photos taken by the UAV during the flight operation are presented. In the Hardware Implementation of UAV & CAV Communication section, the data flow between the CAV and UAV is explained, where DSRC, UDP, TCP and WebSocket communications were used, respectively. In the Use Case Scenario Description section, the use case test scenario is described in detail. In the Test Results section, the success of the CAV and UAV communication is quantified through package drop percentage and latency analysis. In the Conclusion section, conclusions are drawn, and next steps are elaborated.

## II. METHODOLOGY

In this section, each component of the CAV and UAV connectivity hardware implementation is explained. A CAV platform with DSRC connectivity was employed as the ground vehicle. For the aerial vehicle part, rather than choosing an off-the-shelf UAV, individual parts of the UAV were purchased and was put together to get the UAV platform. A Raspberry Pi 4B was mounted on the UAV as a companion computer to handle image processing, connection to CMP and WebSocket servers, as well as data acquisition. In order to have internet connection on the UAV, a 4G HAT was used with Google-Fi service provider to provide 4G internet to the Raspberry Pi 4B, as well as act as a Wi-Fi hotspot to nearby ground crews. A Logitech camera was mounted on the UAV to take pictures and survey the ground while the UAV was flying.

On the cloud side, Contingency Management Platform (CMP) was used to get information from the CAV and UAV and to manage alerts for incidents.

### A. CONNECTED AND AUTOMATED VEHICLE (CAV) PLATFORM

The CAV platform used for this manuscript can be seen below in FIGURE 1. The vehicle is a 2017 model Ford Fusion Hybrid vehicle with Drive-By-Wire. The CAV platform enables testing of V2V, V2X and V2I algorithms, as well as Advanced Driver Assistance systems (ADAS).

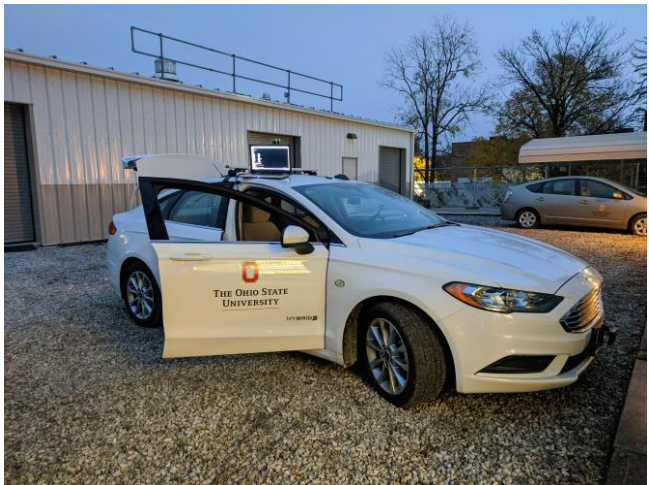

**FIGURE 1.** CAV platform Ford Fusion 2017 vehicle.

For this manuscript, the CAV platform was equipped with a DSRC modem to communicate with a flying UAV, which also was equipped with a DSRC modem.

### B. UNMANNED AERIAL VEHICLE (UAV) PLATFORM

The UAV platform used for this manuscript can be seen below in FIGURE 2. The UAV platform is a hexarotor with 6 rotors and has a 12V voltage regulator. The UAV is equipped with a telematics radio, which enables it to be controlled by a pilot on the ground. The telematics unit also enables the UAV status to be monitored using a ground control station.

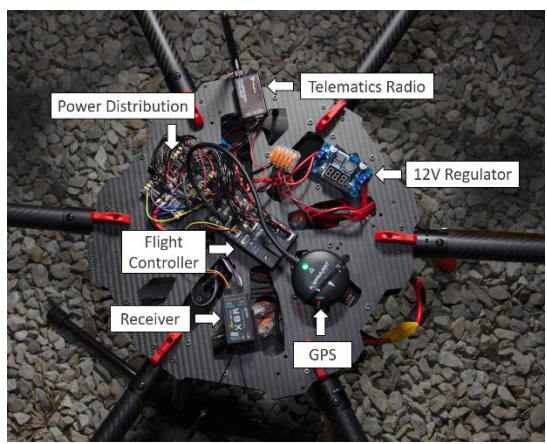

**FIGURE 2.** UAV platform.

For the work presented in this manuscript, the UAV was equipped with a DSRC modem to communicate with the CAV platform. For communication with other system elements, a Raspberry Pi 4B was mounted in the UAV to provide 4G internet. Raspberry Pi 4B was also used to establish communication between the UAV and a WebSocket. Finally, another communication link was established between the UAV and the Contingency Management Platform (CMP).

### C. RASPBERRY PI 4B COMPANION COMPUTER

Raspberry Pi 4B is a low-cost and energy-efficient electronic board mini computer that can be, and has been used, for a variety of tech projects. Raspberry Pi 4B has a 1.5GHz 64-bit quad-core Arm Cortex-A72 CPU, 8 GB RAM, integrated 802.11ac/n wireless LAN, and Bluetooth 5.0 [18]. Additionally, it has 2 USB 2 ports, 2 USB 3 ports and a gigabit ethernet port. The processing power of the Raspberry Pi 4B, as well as its compact size and low weight, makes it an ideal candidate for UAV research and flight operations.

For this study, the Raspberry Pi 4B was chosen as the companion computer and mounted on the UAV platform. It was connected to the camera for image processing. Additionally, Raspberry Pi 4B was connected to the DSRC on-board-unit (OBU) modem on the UAV through ethernet port for a UDP connection. It was also connected through USB connection to the 4G HAT to get 4G internet to the flying UAV. It was also used used for image processing, providing 4G internet to the communication protocols, as well as establishing UDP and TCP communication between different applications and hosts.

### D. 4G INTERNET

Even though DSRC communication is a reliable connection, internet connection to transfer data from the UAV to a server might be necessary in real-life. For remote flights with no Wi-Fi access, equipping a UAV with on-board 4G internet becomes crucial. For that reason, a 4G internet shield was added to the setup for internet connection. By doing so, the WebSocket communication architecture was expanded for more realistic scenarios. For the 4G internet connection, a SIM7600A-H 4G HAT Board was used [19]. A SIM card with T-Mobile network provider was inserted to the card slot and the board was connected to the Raspberry Pi 4B via USB. The necessary libraries and dependencies were resolved.

### E. LOGITECH CAMERA AND IMAGE PROCESSING

Using image processing, it is possible to extract useful information about roadways and vehicles travelling on roadways. CAVs and UAVs can benefit from this information in terms of fuel economy, ride comfort and mobility. For real life testing of CAV and UAV coordination, a live camera link was established between the camera mounted on the UAV platform and the Raspberry Pi 4B companion computer.

For this implementation, a Logitech Webcam was connected through USB to the Raspberry Pi 4B. Python scripts were used along with the OpenCV library image processing tools. Additionally, for display purposes, captured frames from the on-board camera were then sent to the Webserver through Python scripts and WebSocket communication.

## F. WEBSOCKET SERVER

WebSocket servers are applications that are programmed to listen to a TCP connection to ensure full-duplex communication. When WebSocket servers are being programmed, the first step is to make the server listen for incoming socket connections using a standard TCP socket. Then, the handshake must be established, where the details of the connection are negotiated. The WebSocket server also must keep track of the clients which has already completed the handshake. During the connection, either the server or the client can send messages at any time [20].

For this study, the client Raspberry Pi 4B first sent the handshake request to the server. When the connection was established between the WebSocket and the client, then the client Raspberry Pi sent photos taken by the Logitech camera to the WebSocket server. Once the WebSocket server received the photos, the messages and the photos received by the WebSocket server were displayed in a web browser. Other 3rd parties could also see the data received by the WebSocket server by going to the server address in their web browser and completing a handshake with the server as an observer.

## G. CONTINGENCY MANAGEMENT PLATFORM (CMP)

Contingency Management Platform (CMP) is a platform that can detect off-nominal conditions that could affect the UAV operations. Additionally, CMP can provide situational awareness and the impact of the off-nominal condition to UAVs. Expanding on that, it is possible to alert the UAVs to the existence of accidents when they are present. CMP can also be used to alert the UAV about several scenarios, one being flight zones to avoid for the UAV operations. For this real-life implementation, an accident location that was detected by the CAV was sent to the UAV, which then sent it to the CMP to alert all the UAVs connected to the CMP system.

## III. HARDWARE IMPLEMENTATION OF UAV & CAV COMMUNICATION

Using image processing, it is possible to extract useful information about roadways and vehicles travelling on roadways. CAVs and UAVs can benefit from this information in terms of fuel economy, ride comfort and mobility. Since information regarding traffic flow, average vehicle speed, presence of work zones, as well as queues and accidents can be detected using cameras in conjunction with UAVs, a companion computer setup with a camera was prepared.

## A. DSRC COMMUNICATION BETWEEN CAV & UAV

The Dedicated Short Range Communication (DSRC) protocol has been utilized for communication needs in the automotive industry and has been used successfully for a wide variety of applications. Cantas *et al.* [21] and Kavas-

Torris *et al.* [22] utilized DSRC communication for a Vehicle-to-Infrastructure (V2I) algorithm, so that Signal Phasing and Timing (SPaT) messages broadcast by a traffic light could be picked up by a CAV equipped with a DSRC modem to be used for fuel consumption reduction. Gelbal *et al.* [23] designed a Hardware-in-the-Loop (HIL) simulator to test automated driving algorithms and tested a Cooperative Adaptive Cruise Control (CACC) model utilizing DSRC communication for car following applications.

**UAVs can also be equipped with DSRC modems to send data to and receive data from DSRC equipped aerial and ground vehicles. For this test, the CAV platform and the UAV platform were equipped with DENSO WSU 5900 DSRC modems. The modems and antenna configurations on the aerial and ground platform can be seen in**
FIGURE 3.

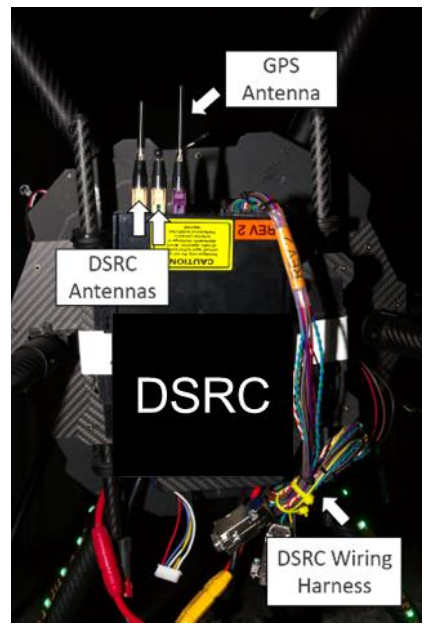

**FIGURE 3.** DSRC modem on the UAV platform.

Preliminary flight testing has been conducted to test the performance of the DSRC communication between a flying UAV and a stationary CAV and results have been published [17]. It has been shown that even though DSRC is a short-range communication protocol, it can be useful in low altitude flight applications.

## B. UDP COMMUNICATION BETWEEN DSRC MODEM AND RASPBERRY PI 4B

The User Datagram Protocol (UDP) is an internet protocol suite that allows computer applications to send messages to other hosts on an Internet Protocol (IP) network. UDP uses a simple connectionless communication mode and provides checksums for data integrity with no error correction facilities. Using UDP, applications can use sockets to establish host-to-host communication. UDP connection prioritizes time over reliability, hence it is faster, but less reliable, than TCP.

A UDP connection was established between the DSRC modem on-board-unit (OBU) and the Raspberry Pi 4B companion computer through the ethernet port. Python libraries were used to open a socket on the DSRC OBU side and receive the data from the open socket on the Raspberry Pi 4B side.

### C. TCP COMMUNICATION BETWEEN RASPBERRY PI 4B AND CMP

The Transmission Control Protocol (TCP) is an internet protocol suite that provides reliable and error-checked delivery of a stream of bytes between applications. Since TCP is a connection-oriented protocol, a connection has to be established between the client and server before data can be sent. Having a TCP connection increases the latency in the data transfer, however, using TCP brings about a 3-way handshake and error detection.

Using the 4G internet on the UAV, a TCP connection was established between the Raspberry Pi 4B companion computer mounted on the UAV and the CMP. At the beginning of the flight test, a System Message with various fields were created to send CAV location information to the CMP continuously. CMP updated the location information sent by the UAV every 1-2 seconds. Then, a Fault Message was created with specific fields to alert the other CMP users about the location of the incident.

### D. WEBSOCKET COMMUNICATION BETWEEN RASPBERRY PI AND WEBSERVER

WebSocket is a computer communication protocol that allows a two-way interactive communication between clients and a server over TCP. Using the WebSocket protocol, the Webserver presented in web server is enabled in the Methodology Section can interact with another web browser or other client applications. Therefore, it is possible to send messages and receive responses between one server and multiple clients. To establish WebSocket communication between a client and a server, an internet connection is required. For this reason, the 4G connection was established as part of the hardware implementation.

Python scripting and WebSocket libraries [24] were used to design an 2-way WebSocket communication portal between the Raspberry Pi 4B and the Webserver. Frames captured from the live feed of the camera were sent to the Webserver and displayed in a web browser.

### E. CONNECTION DIAGRAM FOR THE HARDWARE IMPLEMENTATION

Using the 4G internet on the UAV, a TCP connection was established between the Raspberry Pi 4B companion computer mounted on the UAV and the CMP. At the beginning of the flight test, a System Message with various fields were created to send CAV location information to the CMP continuously. CMP updated the location information sent by the UAV every 1-2 seconds. Then, a Fault Message was created with specific fields to alert the other CMP users about the location of the incident.

The CAV platform equipped with all hardware components can be seen in
**FIGURE 4**.

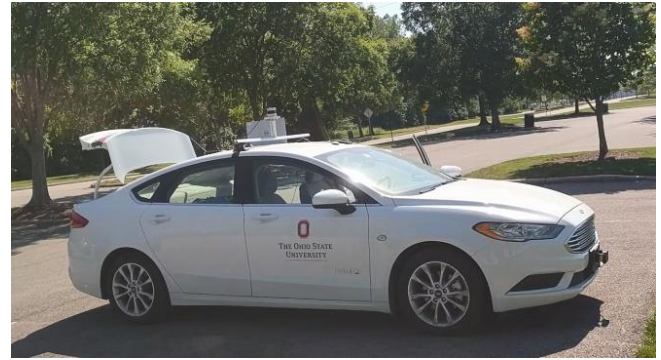

**FIGURE 4.** CAV platform equipped with hardware.

**The UAV platform equipped with all hardware components can be seen in**
FIGURE 5.

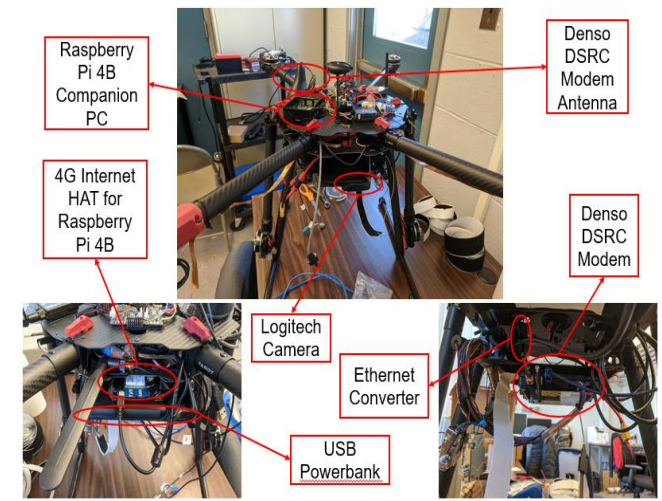

**FIGURE 5.** UAV platform equipped with hardware.

### IV. USE CASE SCENARIO DESCRIPTION

For the experiment, the parameters broadcast from the CAV and received by the UAV through DSRC communication are as follows

- CAV latitude
- CAV longitude
- CAV altitude
- Message Transmit/Receive Timestamp
- GPS Time
- Remote Ground speed (m/s)

For the experiment, the parameters sent to the CMP by the UAV thorough TCP communication in the System Message are as follows
- Message Transmit/Receive Timestamp
- CAV latitude
- CAV longitude
- CAV altitude
- UAV latitude
- UAV longitude
- UAV altitude

For the experiment, the parameters sent to the CMO by the UAV through TCP communication in the Fault Message are as follows
- Incident latitude
- Incident longitude
- Incident altitude
- Incident time
- Incident type

For the use case, there was an active DSRC communication link between the CAV and the UAV, meaning that CAV and UAV shared location information with each other. The CAV location data was then transferred from the UAV OBU to the Raspberry Pi 4B companion PC that was mounted on the UAV through UDP communication. To establish communication between the Raspberry Pi 4B and the Contingency Management Platform (CMP), a TCP connection was set up, where a System Message was generated by the Raspberry Pi 4B. The System Message included information CAV location (latitude, longitude, and altitude) and message transmit time stamp, as well as the UAV location (latitude, longitude and altitude). The System Message was updated with live information from the CAV and UAV on the CMP. When the CAV approached the predetermined accident/incident location, a Fault Message was generated by the Raspberry Pi 4B. The Fault Message included information including incident time, incident type, incident latitude and incident longitude. The Fault Message was sent to the CMP and the incident information was displayed in the CMP.

## V. TEST RESULTS

To test the performance of the system, 2 experiments were run. In the following sections, the results of each experiment are explained in detail. can be grouped under 2 categories.

### A. EXPERIMENT #1 AND RESULTS

For the 1st experiment, DSRC communication was established between the CAV and UAV, so that location information could be shared between the ground and aerial vehicles. UAV telemetry was shared with another PC to monitor the health and state of the UAV by the pilot in QGroundControl. Then, DSRC data was passed to the Raspberry Pi 4B on-board companion PC using UDP to be further processed. Raspberry Pi 4B then established a connection through WebSocket with the WebSocket server, so that vehicle information, as well as photos taken with a camera mounted on the UAV, could be sent to the server to be displayed in a web browser. The schematic for the 1st experiment can be seen in FIGURE 6.

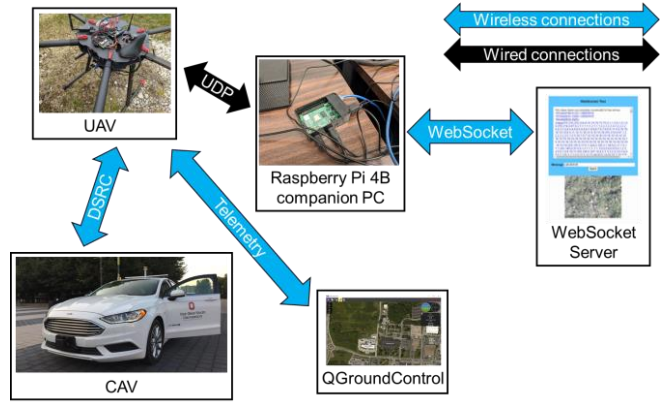

**FIGURE 6.** Schematic for Experiment #1.

Different hardware and software components were used in Experiment #1 to get the data transfer from the CAV OBU all the way to the Raspberry Pi 4B and the WebSocket server. The Raspberry Pi monitoring each communication link during Experiment #1 can be seen below in FIGURE 7.

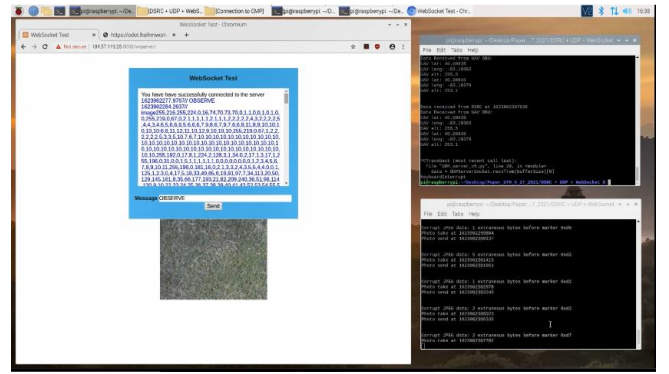

**FIGURE 7.** Raspberry Pi 4B Screen during Experiment #1.

To capture the latency in the system, a detailed latency analysis was carried out for Experiment #1. Latency analysis was done for 3 different cases. For the 1st case, DSRC communication delay between the CAV and the UAV OBU's was calculated, and the average was taken for the data recorded during Experiment #1. For the 2nd case, the UDP communication delay between the UAV OBU and the Raspberry Pi 4B was recorded and the average was taken. Lastly, the WebSocket communication delay between the photo being sent through the socket to the server and the server displaying the photo was recorded and the average was taken. The results are summarized in FIGURE 8. As seen in the figure, the smallest latency was observed for the DSRC communication, followed closely by WebSocket. The slowest communication was observed for the UDP communication.

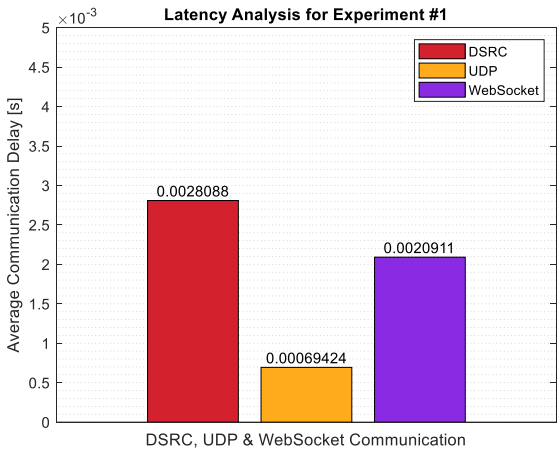

**FIGURE 8.** Latency Analysis for Experiment #1.

Other than the latency analysis, the GPS data from both the CAV and UAV were also plotted for this test. The CAV GPS location (in blue) and UAV GPS position can be seen in FIGURE 9. In the figure, the altitude of the CAV has an increasing trend for some data points due to the drift observed in the GPS.

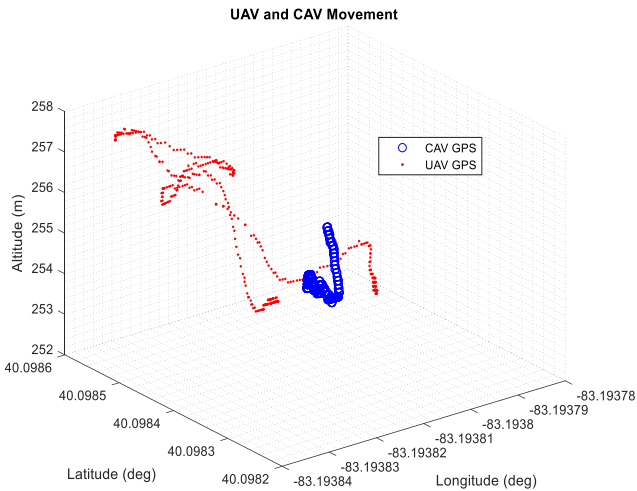

**FIGURE 9.** CAV GPS and UAV GPS Location for Experiment #1.

### B. EXPERIMENT #2 AND RESULTS

The 2nd experiment builds up on the 1st experiment, such that the DSRC and WebSocket connections here are identical. However, in addition to the DSRC and WebSocket communication links in the 1st experiment, the Raspberry Pi 4B companion PC also established a TCP connection to the Contingency Management Platform (CMP) to send vehicle location information and if necessary, accident or incident location to CMP. The schematic for the 2nd experiment can be seen in FIGURE 10.

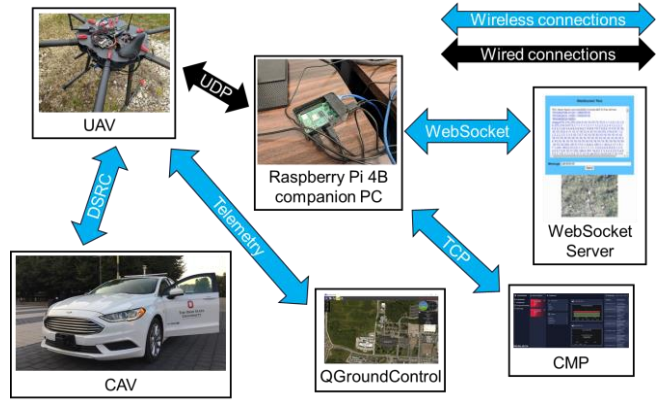

**FIGURE 10.** Schematic for Experiment #2.

Different hardware and software components were used in Experiment #2 to get the data transfer from the CAV OBU all the way to the Raspberry Pi 4B and the WebSocket server. The Raspberry Pi monitoring each communication link during Experiment #2 can be seen below in FIGURE 11.

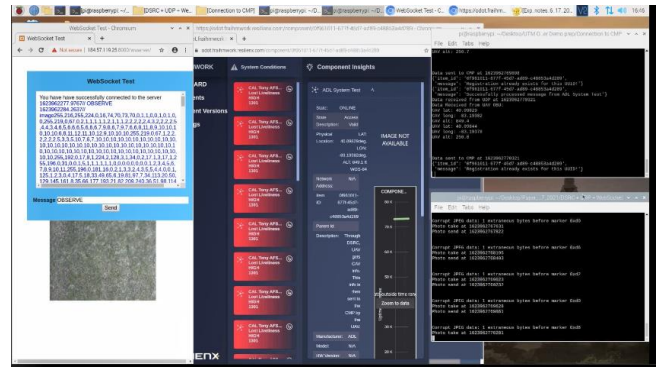

**FIGURE 11.** Raspberry Pi 4B Screen during Experiment #2.

To capture the latency in the system, a detailed latency analysis was carried out for Experiment #2. Latency analysis was done for 4 different cases. For the 1st case, DSRC communication delay between the CAV and the UAV OBU's was calculated, and the average was taken for the data recorded during Experiment #1. For the 2nd case, the UDP communication delay between the UAV OBU and the Raspberry Pi 4B was recorded and the average was taken. For thew 3rd case, the WebSocket communication delay between the photo being sent through the socket to the server and the server displaying the photo was recorded and the average was taken. Lastly, the communication delay between Raspberry Pi 4B and CMP through the TCP connection was recorded, and the average was taken. The results are summarized in FIGURE 12. Looking at FIGURE 12, it is seen that similar to Experiment #1, the smallest latency was observed in DSRC communication. However, the DSRC communication delay in Experiment #2 was larger than that of Experiment #1. The second smallest latency was observed in WebSocket communication, and compared to Experiment #1, the latency was slightly larger in Experiment #2. CMP communication latency through the TCP port between the Raspberry Pi 4B companion computer and the CMP platform was about 0.5 seconds. Lastly, the largest communication delay was observed in the UDP connection, the delay was around 8 seconds, which is considerably larger compared to all other system components, as well as the UDP communication latency in Experiment #1.

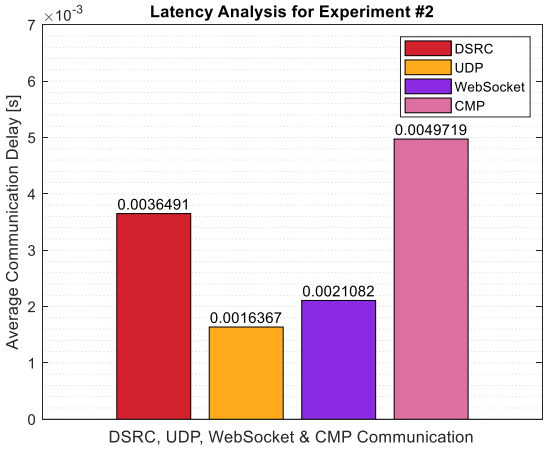

**FIGURE 12.** Latency Analysis for Experiment #2.

Another variation of this test was conducted, where whenever the UAV was flying around, CAV was also in motion and was operated by the driver to drive in a loop around the parking lot. The GPS data from the DSRC communication link between the CAV and UAV were plotted in a 3D plot and can be seen below in FIGURE 13.

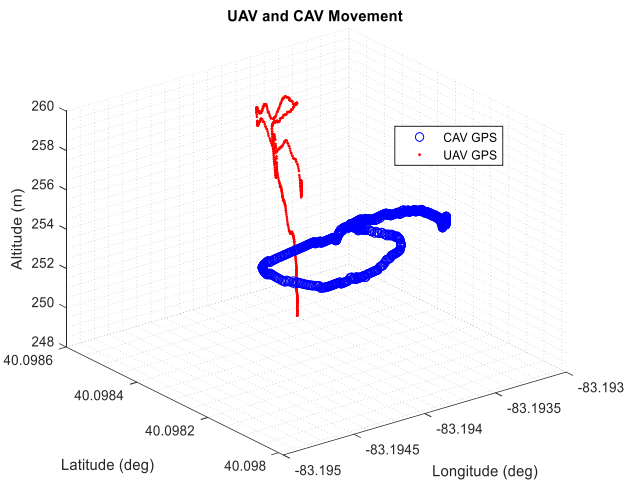

**FIGURE 13.** CAV GPS and UAV GPS Location for Experiment #2.

For the variation of experiment 2 with both CAV and UAV in motion simultaneously, the latency analysis was repeated. The results of this version of the test can be seen in FIGURE 14. Looking at the results, it is seen that the latency in each communication link has increased compared to the version of the Experiment #2 with only UAV in motion with stationary CAV. This result is expected, as during the experiment, the distance between the UAV and CAV changed more rapidly for this variation of the test, and that resulted in a higher latency for the overall system.

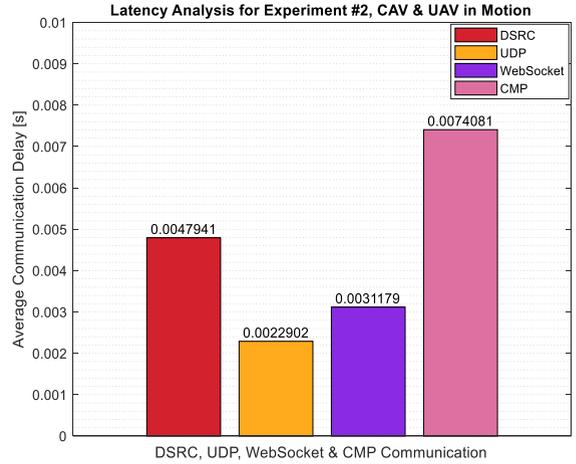

**FIGURE 14.** Latency Analysis for Experiment #2, CAV & UAV in Motion

## VI. CONCLUSION

In this paper, a V2X communication architecture was modeled and tested through real-life testing to show the potential in ground and aerial vehicle communication. Each system component was presented and explained in detail.

Two test scenarios were devised, and those scenarios were tested through real-life testing. The results were analyzed, and special care was given to communication latency for the DSRC communication between the ground and aerial vehicle, UDP communication between the UAV OBU and Raspberry Pi 4B companion PC, WebSocket communication between the Raspberry Pi 4B companion computer and the WebSocket server, and the TCP communication between the Raspberry Pi 4B and CMP. After the latency analysis for Experiment #1, it was seen that the latency was minimal for each component, with DSRC having the least latency and UDP having the largest latency. For Experiment #2, however, it was observed that the UDP communication had a considerably large latency.

For future work, dedicated 4G LTE with priority from service towers could enhance the 4G internet connection, making the connection faster and more reliable, eliminating down time. Control of the CAV and UAV for coordinated and cooperative motion is also of significance for future work. Parameter space based robust control ([25],[26]) and model regulation ([27],[28]) will be a good choices for controlling the path tracking of the CAV ([29]) and UAV as they have successfully been applied before in applications ranging from automotive control ([30],[31],[32],[33]), friction compensation ([34]), robot force control ([35]) to atomic force microscope control ([36]) and collision avoidance ([37]).

## ACKNOWLEDGMENT

The authors thank and acknowledge the support of the Automated Driving Lab at The Ohio State University, The Aerospace Research Center (ARC) at The Ohio State University, and the Ohio Department of Transportation Unmanned Aircraft Traffic Management project. The authors would also like to thank and recognize the support and hardware contributions from DENSO. A DENSO WSU 5900 DSRC modem is being used for physical testing on the automated flight platform.

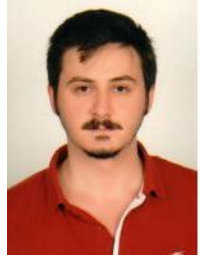

**SUKRU YAREN GELBAL** received his BS in Mechatronics Engineering from Okan University, Istanbul, Turkey in 2014 and MS in Mechatronics Engineering from Istanbul Technical University, Istanbul, Turkey in 2017. He is currently pursuing the PhD in Electrical and Computer Engineering at the Ohio State University, Ohio, USA. His research interests include software and hardware implementation for autonomous driving, HIL simulations, collision avoidance and V2X communication.

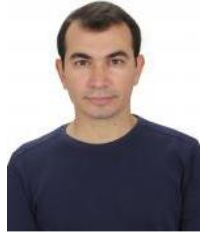

**MUSTAFA RIDVAN CANTAS** received his BS in Electrical and Electronics Engineering from Yeditepe University, Istanbul, Turkey, in 2009, and his M.S. degree in the same field from Bilkent University, Ankara, Turkey, in 2012. Upon graduation, he worked as an Embedded Hardware Design Engineer for OTOKAR Automotive and Defense Industry, Sakarya, Turkey, until 2016. He is currently working toward his PhD in the Automated Driving Lab, Electrical and Computer Engineering, The Ohio State University, OH. His research interests include connected and autonomous vehicles, intelligent transportation systems, advanced driver assistance systems, vehicle control, and fuel economy.

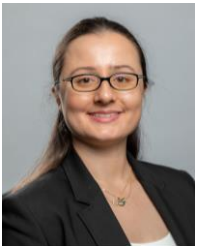

**OZGENUR KAVAS-TORRIS** completed her Bachelor of Science in Mechanical Engineering and Minor in Mechatronics at Middle East Technical University (METU), in 2015, Ankara, Turkey. She completed her Master of Science in Mechanical Engineering at Washington State University, in 2017, Vancouver, USA. She is a PhD. Candidate and is currently working towards her PhD. in Mechanical Engineering at The Ohio State University. She is a member of the Automated Driving Lab (ADL). Her research interests include connected and autonomous vehicle (CAV) systems, CAV control systems for energy management and fuel economy, vehicle speed optimization for fuel economy and mobility, and advanced driver assistance systems (ADAS).

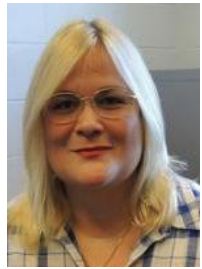

**BILIN AKSUN-GUVENC** is a Research Professor in the Department of Mechanical and Aerospace Engineering at the Ohio State University since Sept. 2017. She received her BS, MS, and PhD in Mechanical Engineering from Istanbul Technical University, Istanbul, Turkey, in 1993, 1996, and 2001, respectively. She worked previously in Istanbul Technical University and Istanbul Okan University where she was a professor. Her expertise is in automotive control systems–primarily vehicle dynamics controllers, such as electronic stability control, adaptive cruise control, cooperative adaptive cruise control, and collision warning and avoidance systems, autonomous vehicles, connected vehicle applications, intelligent transportation systems, smart cities, and smart mobility. She is the co-author of one book, two book chapters and has more than 130 publications in journals and conferences. She is a Member of the International Federation of Automatic Control (IFAC) Committees on Automotive Control and on Mechatronics.

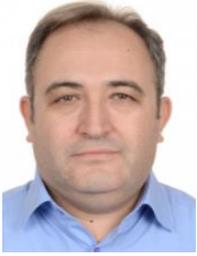 **LEVENT GUVENC** received his PhD in Mechanical Engineering from the Ohio State University, Columbus, OH, USA, in 1992. He is currently a Professor in Mechanical and Aerospace Engineering with the Ohio State University with a joint appointment at the Electrical and Computer Engineering Department. He is a Member of the International Federation of Automatic Control (IFAC) Technical Committees on Automotive Control; Mechatronics; and Intelligent Autonomous Vehicles and the IEEE Technical Committees on Automotive Control; and Intelligent Vehicular Systems and Control. He was the coordinator of team Mekar in the 2011 Grand Cooperative Driving Challenge. He is currently the co-founder and the director of the Automated Driving Laboratory, Ohio State University. He is the co-author of more than 220 technical publications. He is an ASME Fellow.